 \documentclass{article}
\usepackage{graphicx, amsmath}
\usepackage[english]{babel}
\usepackage{color}
\usepackage{graphicx}% Include figure files
\usepackage{dcolumn}% Align table columns on decimal point
\usepackage{bm}% bold math
\usepackage{amsfonts,amssymb,amsmath}

\usepackage[latin1]{inputenc}

\begin{document}

%\preprint{}

\title{Voltage controlled terahertz transmission through GaN quantum wells} %Title of paper



\author{T.~Laurent, R.~Sharma, J. Torres, P. Nouvel, S. Blin, L. Varani, Y. Cordier, M.~Chmielowska, S. Chenot, JP Faurie, B. Beaumont, P. Shiktorov, E. Starikov, V. Gruzinskis, V. Korotyevyev, V. Kochelap}


\date{\today}

\begin{abstract}
We report measurements of radiation transmission in the $0.220$--$0.325$~THz frequency domain through GaN quantum wells grown on sapphire substrates at room and low temperatures. A significant enhancement of the transmitted beam intensity with the applied voltage on the devices under test is found. For a deeper understanding of the physical phenomena involved, these results are compared with a phenomenological theory of light transmission under electric bias relating the transmission enhancement to changes in the differential mobility of the two-dimensional electron gas. 
\end{abstract}

%\pacs{}

%\maketitle must follow title, authors, abstract and \pacs
\maketitle

In the last years, there has been a growing interest in using the terahertz (THz) radiations as a powerful tool to investigate materials/devices/structures in chemistry, biology, physics, medicine and materials science \cite{dragomanQE2004}. To develop the electro-optic systems needed by the THz roadmap in the next years \cite{Tonouchi}, it will become mandatory not only to have efficient emitters and detectors available but also solid-state devices able to control and modify THz beam parameters such as intensity, phase, polarization, direction, etc\dots 
Nowadays GaN quantum wells can be grown with excellent transport properties and a sufficiently high electron-concentration to strongly interact with electromagnetic radiation propagating perpendicularly to the well surface. 
Since transport properties of the 2D electron gas in the quantum well can be easily controlled by an applied voltage, the possibility to achieve a control of the transmitted radiation is expected.   

%Indeed, the THz domain represents a border range of frequencies in the electromagnetic spectrum between microwaves and infrared where intrinsic scientific and technological difficulties appear, leading to several approaches in the realization of THz sources. In the low-frequency part of the THz domain (i.e. f $ \lesssim$ 100 GHz), generation is usually obtained either by starting from various current instabilities appearing in Gunn-, IMPATT-, Schottky-diodes $\dots$ \cite{Gallerano}, which are then placed in a waveguide resonant cavity or by the mixing of CW or pulsed infrared lasers \cite{Ferguson,chimot:193510,mangeney:5551}. From the high-frequency side (i.e.  f $\gtrsim$ 5 THz), generation is usually achieved by optical methods making use of the stimulated radiation coming from quantum cascade lasers \cite{Faist,Koler:2002sw,Scalari,Williams}, p-Ge lasers \cite{Brundermann,Odnoblyudov:2004wj} or gas lasers \cite{Crocker}$\dots$. In each of these ways, the generation efficiency decreases significantly when trying to extend the generation into the central part of the THz region (i.e. 100 GHz $\lesssim$ f $ \lesssim$ 5 THz)\cite{Tonouchi}. 

%Therefore, to develop new and efficient sources, it becomes mandatory to investigate new solid-state structures able to control and eventually amplify the THz radiation.

In this letter we present experimental and theoretical results showing the possibility to control the intensity of THz radiations transmitted through GaN quantum wells by applying low values of dc voltage. 

%\begin{figure}
%%\includegraphics[width=\linewidth]{sample}
%\includegraphics[width=0.9\linewidth]{fig1}
%\caption{\label{sample}Layers description of the studied samples (a), contacts' pattern (b), and conduction band diagram (c). The distances between contacts are $L_l = 39$ and $L_s = 19~\mu$m in the long and short zones, respectively. The horizontal lines of (c) correspond to the three first energy levels of the hetero-layer.}
%\end{figure}
The different layers of the fabricated devices consist of a sapphire substrate, a GaN iron doped template, a GaN $1$~$\mu$m-thick buffer, a $1$~nm-AlN/$21$~nm-Al$_{0.28}$Ga$_{0.72}$N/$3$~nm-GaN active layer covered with Ti/Al ohmic contacts. The layers are grown by molecular beam epitaxy and processed as described in \cite{Cordier}. To couple THz waves with the device, we used interdigitated contacts with variable distances between contacts. Here we present results obtained with devices where the distance between contacts are $L_{s} = 19~\mu$m in the short regions and $L_{l} = 39~\mu$m in the long regions. The width of the device is $W=500~\mu$m.
%The devices are not passivated. Fig.~\ref{sample}(c) shows the conduction band diagram of the structure as a function of depth calculated by solving the Schr\"odinger-Poisson equation. The horizontal lines shows the three first energy levels of the quantum well ($-116$, $-19.6$, and $1.69~$meV compared to the Fermi level) with carriers concentrations of $9.87\times10^{12}$, $1.65\times10^{12}$, and $9.48\times10^{9}~\mbox{cm}^{-2}$. Transmission line models measurements attests of a sheet resistance of $280~\Omega/\square$.
%A scheme of the experimental configuration is sketched in Fig.~\ref{setup}.

The experimental set-up uses a commercial electronic source to generate continuous harmonic waves in the $0.220$ -- $0.325$~THz frequency range. The beam is focused in a He-free cryostat by a first spherical mirror. With the simulation of Gaussian beams propagation in this particular frequency domain, we estimated the position of the source needed to obtain the waist in the center of the cryostat where devices are maintained. Further mechanical alignments are performed to optimize the focusing of the beam on the device. The transmitted beam through device is then extracted from the cryostat and focused with the help of a second spherical mirror on a Si-bolometer for measurement. A lock-in amplifier system is used to extract the part of the bolometer signal corresponding to the transmitted beam.
%\begin{figure}
%\includegraphics[width=\linewidth]{setup}%
%\caption{\label{setup}Experimental configuration for measurements of transmission spectra in the THz frequency domain.}%
%\end{figure}
The experimental protocol is the following: first the device is placed into the cryostat at the position defined by the Gaussian beams simulation, and then thermalized. The passive transmission of the bench without an applied dc-bias to the device is thus measured. Then the same transmission spectrum is acquired applying a dc-bias.  
% the two last measurements give us an idea of the effect of the electric field on the transmission (increasing or decreasing of the signal respectiveley resulting from less or more absorption of the device under the bias effect).In this letter, we will mainly focus on the experimental demonstration of the possible amplification of THz waves by GaN 2-DEG under constant applied bias. Several experiments have been led to investigate the effect of different conditions such as frequency, electric field, polarization, and temperature on the transmission spectra of the GaN quantum wells.]}

\begin{figure}
\includegraphics[width=0.9\linewidth]{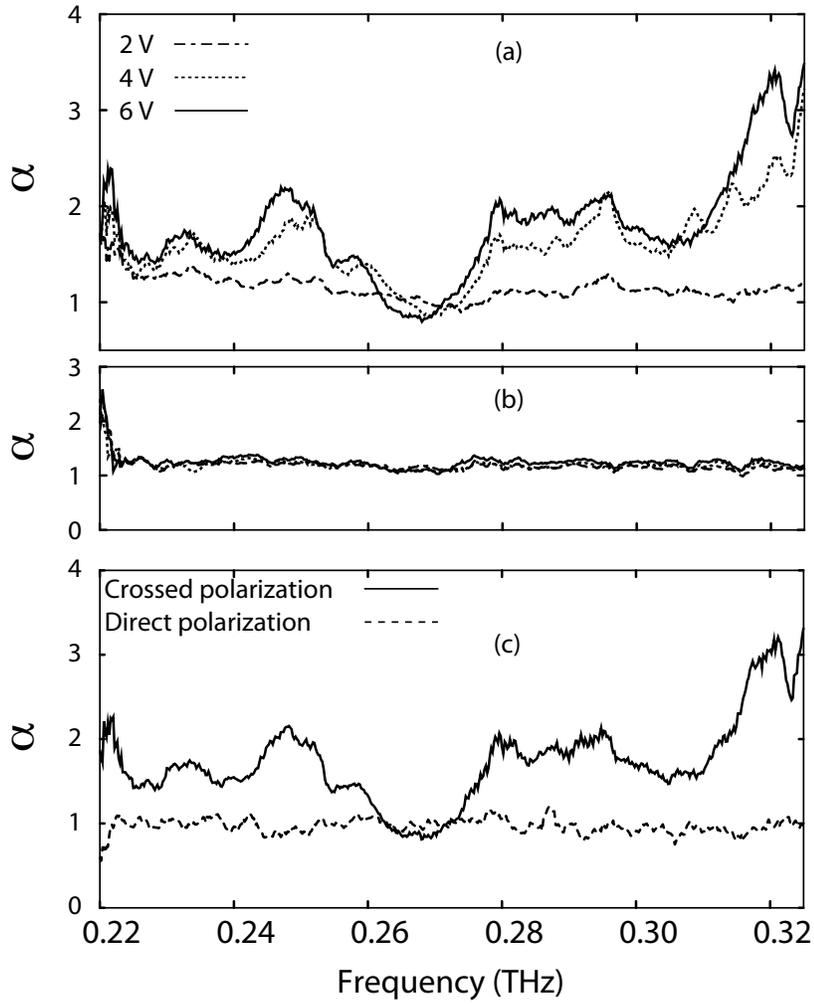}
\caption{\label{spec}Transmission coefficient as a function of frequency, for a device with $L_l = 39$ and $L_s = 19~\mu$m, for three different biases 2 V, 4 V an 6 V and two different temperatures: (a) $77$~K, and (b) $300$~K. (c) Transmission coefficient for two polarizations of the incident beam on the same device at 77 K. 
%Direct polarization : fingers of the samples are parallel to the polarization direction of the incident radiation. Crossed polarization: fingers are perpendicular to the polarization direction. 
}
\end{figure}

The main results are summarized in Fig.~\ref{spec}, showing the transmission coefficient $\alpha$ defined as $ \alpha(f)=T_B(f)/T_0(f)$; where $f$ is the frequency, and $T_B(f)$ and $T_0(f)$ are respectively the measured spectra with a bias $B$ and without bias applied to the device. The applied voltages of investigation vary between $0$ and $6$~V which correspond to average electric fields in the conduction channel from 0 up to $0.103$ and $0.21$~kV/cm in the long and short regions of the device respectively. On Fig.~\ref{spec}(a), for measurements performed with the samples maintained at $77$~K, a strong enhancement of the transmitted beam increasing the voltage from $0$ to $6$~V is found practically in the whole investigated frequency domain. 
On the opposite, on Fig.~\ref{spec}(b), corresponding to room temperature measurements, no significant dependency with the applied voltage is measured. Moreover, the average value of $\alpha$ at $300$~K is near its value at $77$~K when $2$~V is applied, i.e., when no significant enhancement is measured. %This result is in qualitative agreement with Monte-Carlo simulations~\cite{Starikov2001} where the effect of the temperature increases the collisions of electrons with acoustical phonons, and by consequence, destroys the phenomenon of amplification. 

%(d) Effect of the geometrical parameters of the device. $L_s$ varies from $ 6~\mu$m to $19~\mu$m while $L_l$ varies from $52~\mu$m to $39~\mu$m. (c) and (d) were obtained at $77$~K.
It is noteworthy to emphasize that $\alpha$ depends on the orientation of the sample with respect to the polarization direction of the incident radiation. As shown on Fig. \ref{spec}(c), if the fingers of the samples are parallel to the polarization direction (i.e. direct polarization), no effect of the bias or temperature (not shown) is found, whereas if the fingers are perpendicular to the polarization direction (i.e. crossed polarization), a maximum effect is found. This situation corresponds to the results presented in Fig.~\ref{spec}(a) and (b).

To interpret on a physical basis the phenomena responsible for the transmission enhancement in the sub-THz frequency range 
we have developed a phenomenological theory of light transmission via device under electrical bias that will be published elsewhere\cite{Korotyeyev}. 
We modeled its complicated layered structure 
as a $\delta$-like heavily-doped GaN active layer described by a 2D conductivity $\sigma$ and a sapphire substrate of  
thickness $h$ and dielectric constant $\epsilon$. So far as the thicknesses of other undoped layers are much smaller than the sub-THz wavelengths, they can be ignored. We assumed that $\sigma$ depends on both frequency $f$ and applied voltage $U$ according to the Drude-Lorentz model:  
$\sigma_{f,U}=en^{2D}\mu_{U}/(1-i2\pi fm^{*}\mu_{U}/e)$, where $e$ is the elementary charge, $n^{2D}$ the 2D electron concentration, $m^{*}$  the GaN effective mass  and $\mu_{U}$ the voltage-dependent differential electron mobility. In the framework of this model, analytical expressions for the transmission coefficient  
$\alpha$ can be obtained following the standard procedure as described in ref.~\cite{Born}
%\begin{widetext}
%\begin{equation}
%\alpha(f)=\frac{\left(1+\frac{\Gamma_{f,U}^{\prime}}{2}\right)^{2}+
%\frac{\Gamma_{f,U}^{\prime\prime 2}}{4}+
%\frac{\epsilon-1}{4\epsilon}
%\left[ (\epsilon-(1+\Gamma_{f,U}^{\prime})^{2}-
%\Gamma_{f,U}^{\prime\prime 2})\sin^{2}(k_{2}h)
%-\sqrt{\epsilon}\Gamma_{f,U}^{\prime\prime}\sin(2k_{2}h)\right]}
%{\left(1+\frac{\Gamma_{f,0}^{\prime}}{2}\right)^{2}+
%\frac{\Gamma_{f,0}^{\prime\prime 2}}{4}+
%\frac{\epsilon-1}{4\epsilon}
%\left[ (\epsilon-(1+\Gamma_{f,0}^{\prime})^{2}-
%\Gamma_{f,0}^{\prime\prime 2})\sin^{2}(k_{2}h)
%-\sqrt{\epsilon}\Gamma_{f,0}^{\prime\prime}\sin(2k_{2}h)\right]},
%\label{alpha} 
%\end{equation}
%\end{widetext}
%where $\Gamma^{\prime,\prime\prime}_{f,U}=4\pi/c Re[\sigma(f,U)],Im[\sigma(f,U)]$, %$k_{2}=2\pi f\sqrt{\epsilon}/c$, 
%c is light velocity. 

\begin{figure}
\includegraphics[width=0.9\linewidth]{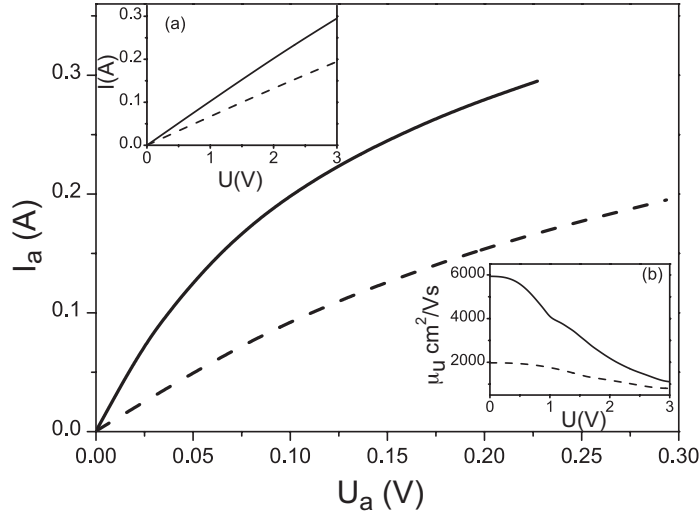}
\caption{Current-voltage characteristics related to the active layer of 2D electron gas. Solid and dashed curves correspond to 77 and 300 K, respectively. Insert: (a) 
Current-voltage characteristics of the whole device and (b) differential mobility as a function of the total applied voltage.}
\label{CV}
\end{figure}

\begin{figure}
\includegraphics[width=0.9\linewidth]{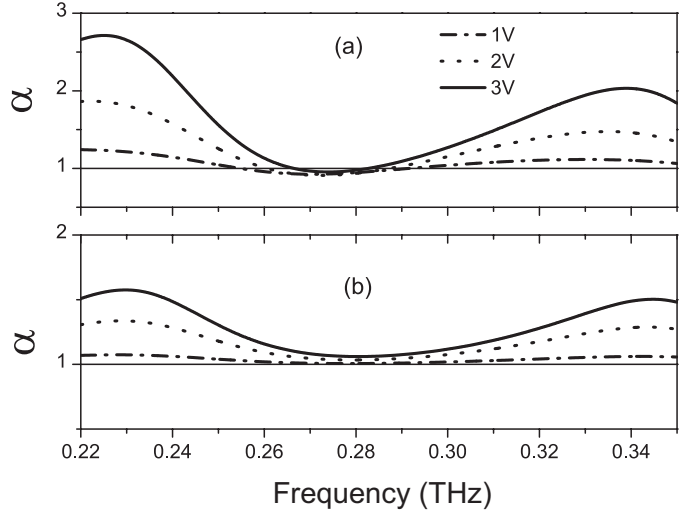}
\caption{Theoretically calculated transmission coefficient as a function of frequency, for $U=1, 2, 3$~V at (a) $77$~K and (b) $300$~K. Calculations are performed assuming  a sapphire substrate width $h=350$~$\mu$m and a dielectric constant $\epsilon=12$. }
\label{alpha_fig}
\end{figure}

The voltage dependence of $\mu_{U}$ can be extracted from the electrical measurements
of current-voltage characteristics $I(U)$. From one hand, using the measured $I(U)$ (see insert (a) of Fig.~\ref{CV}), one can estimate 
a device global resistance $R=15$ $\Omega$ and $R=9.73$ $\Omega$ at room and nitrogen temperatures, respectively.
From the other hand, the active-area resistance $R_{a}$ can be expressed as $R_{a}=[1/(en^{2D}\mu_{0}W)][L_{l}L_{s}/(L_{l}+L_{s})]/N$ (short and long regions
in-parallel), $N\sim \mathrm{Int}[W/(L_{l}+L_{s})]=8$. Capacitance voltage and Hall measurements attest the presence of a 2D electron gas with concentration $n^{2D}\simeq 10^{13}$~cm$^{-2}$ and ohmic mobility 
$\mu_{0}\simeq 2000$~cm$^{2}$/(Vs) at room temperature. For liquid nitrogen temperature we used $\mu_{0}\simeq 6000$~cm$^{2}$/Vs according to Ref.~\cite{Brenon}.
As a consequence, the active-area resistance $R_a$ and the contact resistance $R_{c}=R-R_{a}$ are respectively of $1$ and $14$~$\Omega$ at 300~K 
and respectively of $0.33$ and $9.4$~$\Omega$ at 77~K. The current-voltage characteristics $I_{a}(U_{a})$ of the 2D electron gas shown in 
Fig.~\ref{CV} have been extracted using the relations: $U_{a}=U-IR_{c}$, $I_{a}=I$. 
At 300~K the current $I_{a}$ exhibits only a weak nonlinearity in the whole interval of applied voltages while at 77~K it shows a much more pronounced sub-linear behavior. Such behaviors can be explained  by two possibilities: either a pure electron  
heating or a Joule heating of the whole structure~\cite{Sokolov}. However, the former possibility should be excluded since it is realized in GaN at fields above few kV/cm~\cite{Eastman,starikov:083701} while in our case the applied electric fields are of few hundreds of V/cm. A more accurate current-voltage characteristics of the active layer can be obtained taking into account the details of the heat sink and the temperature dependence of the contact resistance. However, at the present stage of investigations, such a task is beyond the scope of our qualitative theory. From the current-voltage characteristics $I_{a}(U_{a})$ we calculated the differential resistance of the active region, then the mobility $\mu_{U}$ (insert (b) of Fig.~\ref{CV}) and finally the transmission coefficient $\alpha$ shown in Fig.~\ref{alpha_fig}. 
As seen from Fig.~\ref{spec} and~\ref{alpha_fig} the theoretical and experimental results are in good agreement. At 300~K both experimental and theoretical 
$\alpha$ have values in the range $1-1.3$ in the whole frequency domain and a slight increase with $U$ which is attributed to the weak changes of the 
differential mobility $\mu_{U}$ with the applied voltage. At $77$~K,  $\mu_{U}$ decreases by a factor of 6 with increasing bias from 0 to 3~V and therefore, a strong enhancement of the transmission is observed. It should be noted that both experimental and theoretical curves show a frequency window near 0.27~THz where $\alpha$ has a minimum with values around 1 or slightly less then 1. By using the theoretical model, this frequency range is clearly related to radiation interference on the substrate.

\begin{figure}
\includegraphics[width=0.9\linewidth]{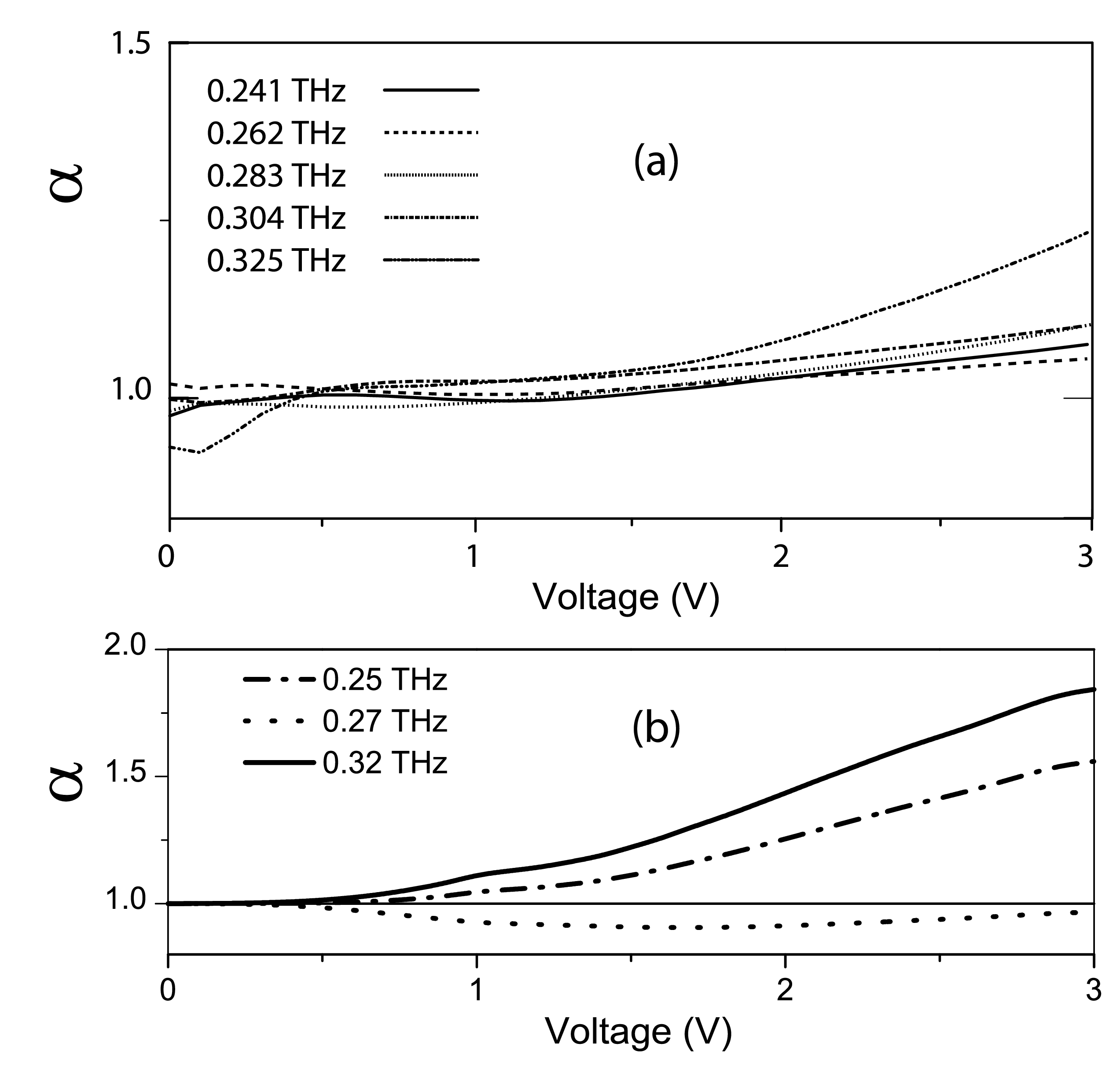}
\caption{Experimental (a) and theoretical (b)  
transmission coefficients as functions of applied voltage for 
three different frequencies $f=0.25, 0.27, 0.32$~THz at 77~K.}
\label{alpha_U}
\end{figure}

The transmission coefficient as a function of applied voltage exhibits different behaviors according to the considered frequency as reported in fig.~\ref{alpha_U}. At frequencies outside the interference region the trend of all the curves is to increase with applied voltage starting from 1~V. Within the interference region the curves show a slight decrease with the voltage.

In conclusion, an original experimental configuration has allowed measurements of the transmitted signal intensity through GaN quantum wells in the 
$0.220$--$0.325$~THz frequency range. Different effects on the transmission spectra such as bias conditions and temperature of the device have been investigated and the measured amplification of the transmitted THz radiation has been theoretically related to a decrease of the differential mobility with the voltage. These results can be considered a promising step toward the development of efficient electro-optic modulators operating in the THz domain. 
The French National Research Agency (ANR) is acknowledged for funding this research under contract AITHER no. ANR-07-BLAN-0321. Moreover we acknowledge also the support of TeraLab-Montpellier (GIS CNRS). 
%\end{acknowledgments}

%\bibliography{Laurent.bib}

\end{document}